\documentclass[preprint, authoryear, 12pt]{elsarticle}
\usepackage{graphicx}
\usepackage{natbib}
\usepackage{amssymb}
\usepackage{subfigure}
\usepackage{color}
\usepackage{algorithm}
\usepackage{algpseudocode}

\begin{document}

\begin{frontmatter}

\title{A hybrid RANS/LES framework to investigate spatially developing turbulent boundary layers}

\author[cornell]{Sunil K. Arolla \corref{cor1}\fnref{fn1}}
\ead{ska62@cornell.edu}

\address[cornell]{Sibley School of Mechanical and Aerospace Engineering, Cornell University, Ithaca, NY 14853, USA}

\cortext[cor1]{Corresponding author}
\fntext[fn1]{Postdoctoral research associate}

\begin{abstract}
A hybrid RANS/LES framework is developed based on a recently proposed Improved Delayed Detached Eddy Simulation (IDDES) model combined with a variant of recycling and rescaling method of generating inflow turbulence. This framework was applied to investigate spatially developing flat plate turbulent boundary layer up to momentum thickness Reynolds number, $R_{\theta}=31000$ and the results are compared with the available experimental data. Good agreement was obtained for the global quantities such as mean velocity and skin friction at all momentum thickness Reynolds numbers considered. The trends obtained for the Reynolds stress components are in the right direction. At high $R_{\theta}$, the shear stress distribution shows significant differences close to the wall indicating scope for further improving the near-wall modeling in such methods.
\end{abstract}


\end{frontmatter}

\section{Introduction}\label{sec:intro}
In the industrial CFD, there is an increasing trend for using eddy-resolving methods, where at least a part of the turbulence spectrum is resolved in at least a part of the flow configuration. For high Reynolds number wall-bounded flows such as those over aircrafts' wings, ships' hulls, and in the atmospheric turbulent boundary layer, the use of large eddy simulation with near-wall resolution (LES-NWR) is infeasible owing to the expensive grid resolution requirements. So, hybrid RANS/LES and wall-modeled LES approaches have become models of choice for such applications. The essential idea is to employ RANS model in the near-wall attached boundary layer region (or a wall model) and LES away from the surface. This reduces the computational cost significantly. There has been intensive research in this area and several variants of such models have been proposed in the recent years \citep{frohlich:2008,shur:2008,girimaji:2006,wang:2002,marusic:2010,gritskevich:2012}. However, to establish confidence in using such methods for practical design, rigorous testing over a wide range of flows need to performed. Spatially developing turbulent boundary layer is one such benchmark problem to evaluate and improve the models.

The particular type of hybrid RANS/LES formulation chosen in this work belongs to the Detached Eddy Simulation category. In this method, switch from RANS to LES takes place depending on the local grid spacing and turbulence length scale. This approach relies on inherent flow instability for a quick generation of turbulent content in the LES mode. Due to the explicit dependence on the grid resolution, DES could result in grid induced separation if the grid is not chosen carefully. To address this issue, a shielding function is introduced to prevent the model from switching to LES within the near-wall attached boundary layer region \citep{menter:2002,spalart:2006a}. This approach is called Delayed Detached Eddy Simulation. The original intent of DES is to use it for massively separated flow regions, but it has an ability to function as a type of wall-modeled LES. Initial attempts made to use this formulation for channel flow resulted in an another issue called log-layer mismatch \citep{nikitin:2000,piomelli:2003}. To address this problem, \cite{shur:2008} proposed a modification of the length scale used in the LES region that gives Improved Delayed Detached Eddy Simulation (IDDES) formulation. \cite{gritskevich:2012} fine-tuned these approaches for $k-\omega$ SST as a background RANS model. This IDDES model can function similar to a wall-modeled LES approach for the attached boundary layers. Prior to the current work, IDDES has not been tested on high Reynolds number flat plate turbulent boundary layers rigorously.

While the experiments on turbulent boundary layers reached up to $R_{\theta}=31000$ \citep{degraaff:2000}, direct numerical simulations of the spatially developing, smooth-wall turbulent boundary layer has hitherto been limited to $R_{\theta}=3-5\times10^{3}$. Wall-resolved LES has been shown to work well at moderately high Reynolds numbers \citep{schlatter:2010}, but the cost scales approximately as Reynolds number to the power 1.8 \citep{pope:2000} and hence is not feasible for using at large $R_{\theta}$. Recently, \cite{inoue:2011} performed wall-modeled LES simulations of zero-pressure-gradient flat plate turbulent boundary layer simulations up to $R_{\theta}=O(10^{12})$. They concluded that wall-modeled LES can be used to predict useful properties such as skin friction and mean velocity profiles without having to use fine grid resolution. Their simulations displayed some near-wall issues which they attribute to the wall modeling and the choice of grid resolution. Their work forms the motivation for investigating the accuracy of a hybrid RANS/LES type of approach for flat plate turbulent boundary layer at high momentum thickness Reynolds numbers.

In this work, a variant of recycling and rescaling method of generating inflow turbulence \citep{arolla:2014} is used in combination with $k-\omega$ SST based IDDES approach. This framework is applied to investigate spatially developing flat plate turbulent boundary layers up to $R_{\theta}=31000$. The objective is to understand the applicability and scope of the IDDES approach for using at high Reynolds number attached boundary layers. The hybrid RANS/LES framework presented in this work can also be used for investigating spatially developing turbulent boundary layers in the presence of pressure gradients.

\section{Hybrid RANS/LES framework}
The computational framework used in this research is that of OpenFOAM finite volume based incompressible flow solver. The governing equations solved are:

\begin{eqnarray}
\partial_{i}\hat{u_{i}} &=& 0 \\
\partial_{t}\hat{u_{i}}+\partial_{j}\hat{u_{j}}\hat{u_{i}} &=& -\frac{1}{\rho}\partial_{i}\hat{p} + \nu \nabla^{2}\hat{u_{i}} - \partial_{i}\tau_{ij}
\end{eqnarray}
where $\hat{u_{i}}$ is the mean velocity field. The unclosed term that arises due to averaging is closed by invoking an IDDES model. As an example for hybrid RANS/LES approaches, a recently proposed simplified version of IDDES for $k-\omega$ SST model \citep{gritskevich:2012} has been implemented within OpenFOAM framework. Stated briefly, the transport equations for turbulent kinetic energy ($k$) and specific dissipation rate ($\omega$) are of the following form:

\begin{eqnarray}
\frac{\partial k}{\partial t} + u_{j} \frac{\partial k}{\partial x_{j}} = P_{k}- \sqrt{k^{3}}/l_{IDDES} + \frac{\partial}{\partial x_{j}}\left[\left(\nu+\frac{\nu_{T}}{\sigma_{k}} \right) \frac{\partial k}{\partial x_{j}} \right] \\
\newline
\frac{\partial \omega}{\partial t} + u_{j} \frac{\partial \omega}{\partial x_{j}} = P_{\omega}- D_{\omega} + CD_{\omega} + \frac{\partial}{\partial x_{j}}\left[\left(\nu+\frac{\nu_{T}}{\sigma_{\omega}} \right) \frac{\partial \omega}{\partial x_{j}} \right]
\end{eqnarray}
where $P_{k}$ is the production of turbulent kinetic energy, $P_{\omega}$ is the production of specific dissipation rate, $D_{\omega}$ is the dissipation of $\omega$, and $CD_{\omega}$ is the cross-diffusion term in $\omega$. The term $l_{IDDES}$ is the length scale that operates the switch between RANS and LES. The eddy viscosity is of the form $\nu_{T}={k}/{\omega}$ with a limiter for separated flows. For detailed description of these terms and empirical constants, see \cite{gritskevich:2012}. The governing equations solved are similar to that of LES, but sub-grid scale stress term is replaced by a modeled Reynolds stress.

For the numerical simulations presented in this article, Pressure Implicit with Splitting of Operator (PISO) algorithm is employed. A second order accurate backward implicit scheme for time discretization and a second order central scheme (with filtering for high-frequency ringing) for spatial discretization is used.

\subsection{A variant of recycling and rescaling method for inflow turbulence generation}
The essential idea presented in this paper is based on the work by \cite{spalart:2006} to simplify the inflow generation algorithm using the following physical arguments:
\begin{itemize}
\item The near-wall turbulence regenerates itself much faster than the outer region turbulence $\to$ Apply outer layer scaling throughout.
\item When the recycling station is located quite close to the inflow, which is desirable in terms of computing cost, the conflict between inner and outer region scaling essentially vanishes $\to$ Short recycling distance
\item Corrections to the wall normal velocity component $v$ have very little effect $\to$ Omitted
\end{itemize}

\begin{figure}[!htp]
\centering{
\includegraphics[width=80mm]{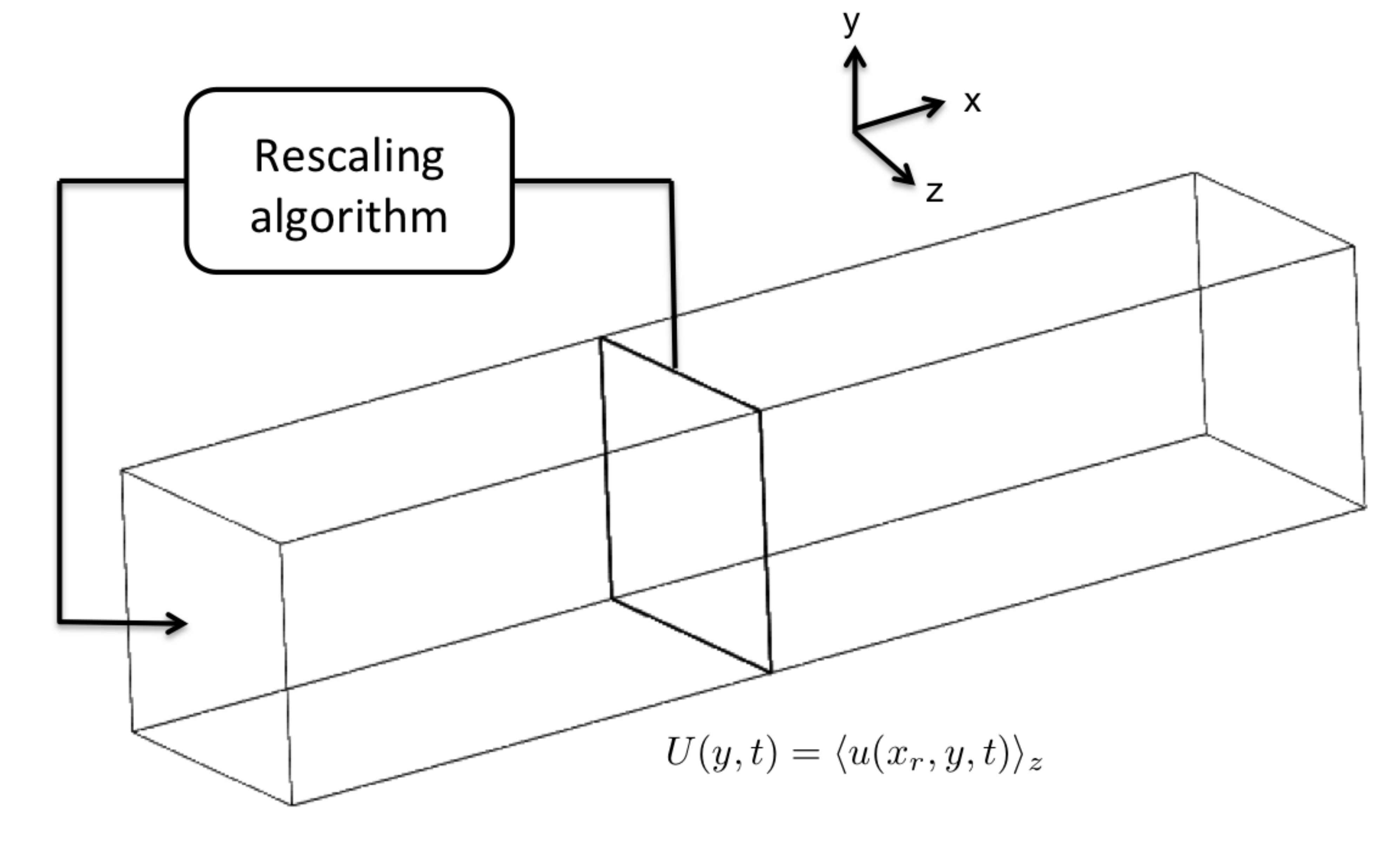}
}
\caption{A schematic of the computational domain used for flat plate turbulent boundary layer simulation. The recycling plane is located at $x_{r}=5 \delta_{0}$ from the inflow boundary.}
\label{flatplate_config}
\end{figure}

In the current work, momentum thickness based scaling is used in place of $99\%$ boundary layer thickness. This avoids the need of locating the edge of the boundary layer. Moreover, using integral quantities like momentum thickness (or displacement thickness) is numerically robust. Most experiments report the momentum thickness Reynolds number at the inflow and hence back-to-back simulations can be set-up easily. A spanwise mirroring method \citep{jewkes:2011} is adopted as it was found to be adequate in the current work for disorganizing unphysical spanwise durable structures.

The steps involved in the inflow generation algorithm are:
\begin{enumerate}
\item Extract the velocity field, $u(x_{r},y,z,t)$, at the recycling plane located at $x_{r}$ and project on to the inflow boundary (see figure \ref{flatplate_config}).
\item Perform spanwise averaging to get $U(y,t) = \langle u(x,y,t) \rangle_{z}$. A simple indexing algorithm is used for the averaging. It involves looping over all the faces and index faces with the same wall-normal coordinate. Since the recycling plane is fixed, this indexing can be stored at the preprocessing step itself and reused at each timestep.
\item Find the freestream velocity $U_{\infty}=U(y_{max},t)$.
\item Integrate the velocity profile to compute the momentum thickness:
\begin{equation}
  \theta_{r} = \int_{0}^{y_{max}}\frac{U(y,t)}{U_{\infty}}\left(1-\frac{U(y,t)}{U_{\infty}} \right) dy
\end{equation}
\item Compute the rescaling factor, $\gamma = \theta_{r}/\theta_{in}$, where $\theta_{in}$ is the desired momentum thickness at the inflow.
\item Rescale only the x-component of the velocity field:
\begin{equation}
  u(x_{in},y,z,t) = u(x_{r},y\gamma,z,t-\Delta t)
\end{equation}
For the Improved Delayed Detached Eddy Simulation (IDDES) model considered in this work, the rescaling operation on the underlying SST variant of $k-\omega$ model requires the following:

\begin{equation}
  k(x_{in},y,z,t) = k(x_{r},y\gamma,z,t-\Delta t)
\end{equation}

\begin{equation}
  \omega(x_{in},y,z,t) = \omega(x_{r},y\gamma,z,t-\Delta t)
\end{equation}
where $t-\Delta t$ means the velocity from the previous time step is used for convenience. A linear interpolation is used to compute velocity at the rescaled y-coordinate.

\item Apply spanwise mirroring to disorganize unphysical structures which would otherwise be recycled and take much time to get dampened by the spanwise diffusion.

\begin{eqnarray}
  u(x_{in},y,z,t) = u(x_{in},y,\Delta z-z,t)  \nonumber \\
  v(x_{in},y,z,t) = v(x_{in},y,\Delta z - z,t) \nonumber  \\
  w(x_{in},y,z,t) = -w(x_{in},y,\Delta z - z,t)
\end{eqnarray}
where $\Delta z$ is considered to be equal to the spanwise period. Note that $w$ has to be negative to ensure spatial coherence once mirrored \cite{jewkes:2011}.

\item Check for constant mass flow rate at the inflow by verifying the bulk velocity.
\end{enumerate}

This framework can be used for investigating spatially developing turbulent boundary layers. The current focus is on zero-pressure gradient flat plate problem.

\section{Flow configuration}
The flow configuration considered is that of a flat plate turbulent boundary layer with inflow momentum thickness Reynolds number in the range, $R_{\theta}=1520-31000$. The computational domain has dimensions $12\delta_{0} \times 3\delta_{0} \times 3\delta_{0}$ in the streamwise, wall-normal, and spanwise directions, respectively where $\delta_{0}$ is the $99 \%$ boundary layer thickness at the recycling plane. The mesh resolution is chosen based on the wall-modeled LES simulations of \cite{inoue:2011}, as given in table \ref{tab:parameters}. Uniform mesh is used in the streamwise and spanwise directions while a hyperbolic tangent stretching is used in the wall-normal direction to cluster points close to the wall. The recycling station was located at $5 \delta_{0}$ downstream of the inlet and the simulation provides its own inflow. The bottom wall is treated as a no-slip wall, top boundary is a slip wall, and at the outflow an advective boundary condition is used. Initial conditions are provided using a mean velocity profile given by Spalding law with random fluctuations with a maximum amplitude of $10\%$ of the freestream value superimposed on the mean value. The time step used is approximately two viscous time units ($\Delta t \approx 2\nu/u_{\tau}^{2}$). The simulation was run for 1000 inertial timescales ($\delta_{0}/U_{\infty}$) to eliminate transients and the statistics are collected over another 1000 timescales.

\begin{table}[!htp]
  \begin{center}
\def~{\hphantom{0}}
  \begin{tabular}{c c c c c c c c}
\hline 
       Case &  $R_{\theta}$  &  $N_{x}$ & $N_{y}$ & $N_{z}$ & $\Delta_{x}^{+}$ & ${\Delta_{y}^{+}}_{max}$ & $\Delta_{z}^{+}$ \\
        A    &  1520   &   140   &    96    &   116    &    60     &   20    &   15      \\
        B    &  5200   &   240   &    128   &   128    &    104    &   42    &   42      \\
        C    &  13000  &   240   &    128   &   128    &    245    &   98    &   98      \\
        D    &  31000  &   240   &    128   &   128    &    546    &   219   &   219     \\
\hline
  \end{tabular}
  \caption{Simulation parameters}
  \label{tab:parameters}
  \end{center}
\end{table}

\section{Results and discussion}
\begin{figure}[!htp]
\begin{minipage}[t!]{80mm}
\centering{
\subfigure[Case A: $R_{\theta}=1520$]{\includegraphics[width=80mm]{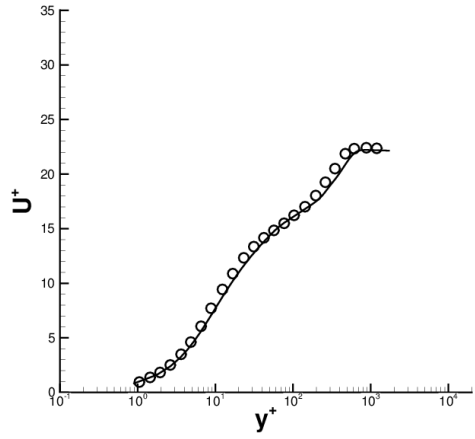}}
}
\end{minipage}
\hfill
\begin{minipage}[t!]{80mm}
\centering{
\subfigure[Case B: $R_{\theta}=5200$]{\includegraphics[width=80mm]{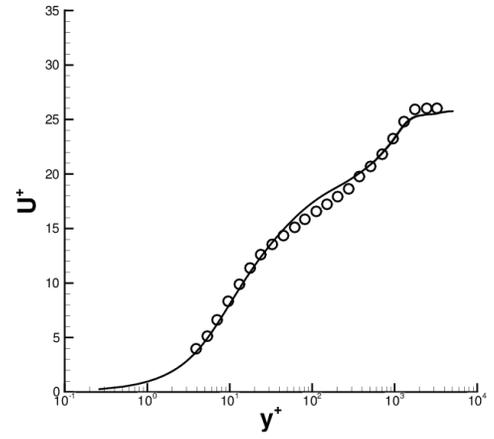}}
}
\end{minipage}
\hfill
\begin{minipage}[t!]{80mm}
\centering{
\subfigure[Case C: $R_{\theta}=13000$]{\includegraphics[width=80mm]{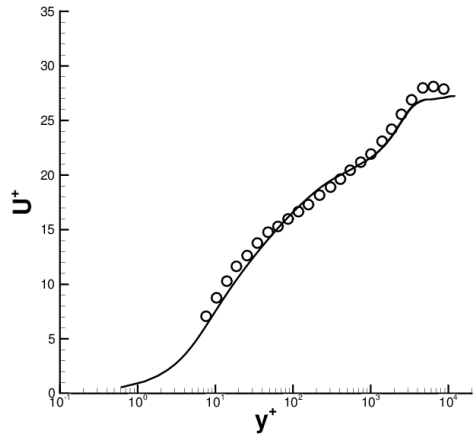}}
}
\end{minipage}
\hfill
\begin{minipage}[t!]{80mm}
\centering{
\subfigure[Case D: $R_{\theta}=31000$]{\includegraphics[width=80mm]{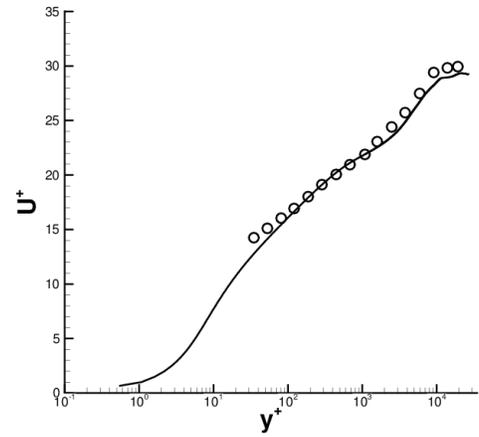}}
}
\end{minipage}
\caption{Mean velocity profiles in comparison with the experimental data. Solid lines: IDDES; Symbols: \cite{degraaff:2000}}
\label{Umean}
\end{figure}

Mean velocity profiles plotted in inner-inner coordinates at different $R_{\theta}$ are presented in figure \ref{Umean}. Comparison with the experimental data of \cite{degraaff:2000} shows that good agreement is obtained for all the cases considered. The mean velocity at the edge of the boundary layer normalized by the friction velocity, $U_{e}^{+}=U_{\infty}/u_{\tau}$, increases with the increase in $R_{\theta}$. The velocity profiles scale well when plotted in the inner coordinates, which is in agreement with the experimental observations.

\begin{figure}[!htp]
\begin{minipage}[t!]{80mm}
\centering{
\subfigure[Streamwise component]{\includegraphics[width=80mm]{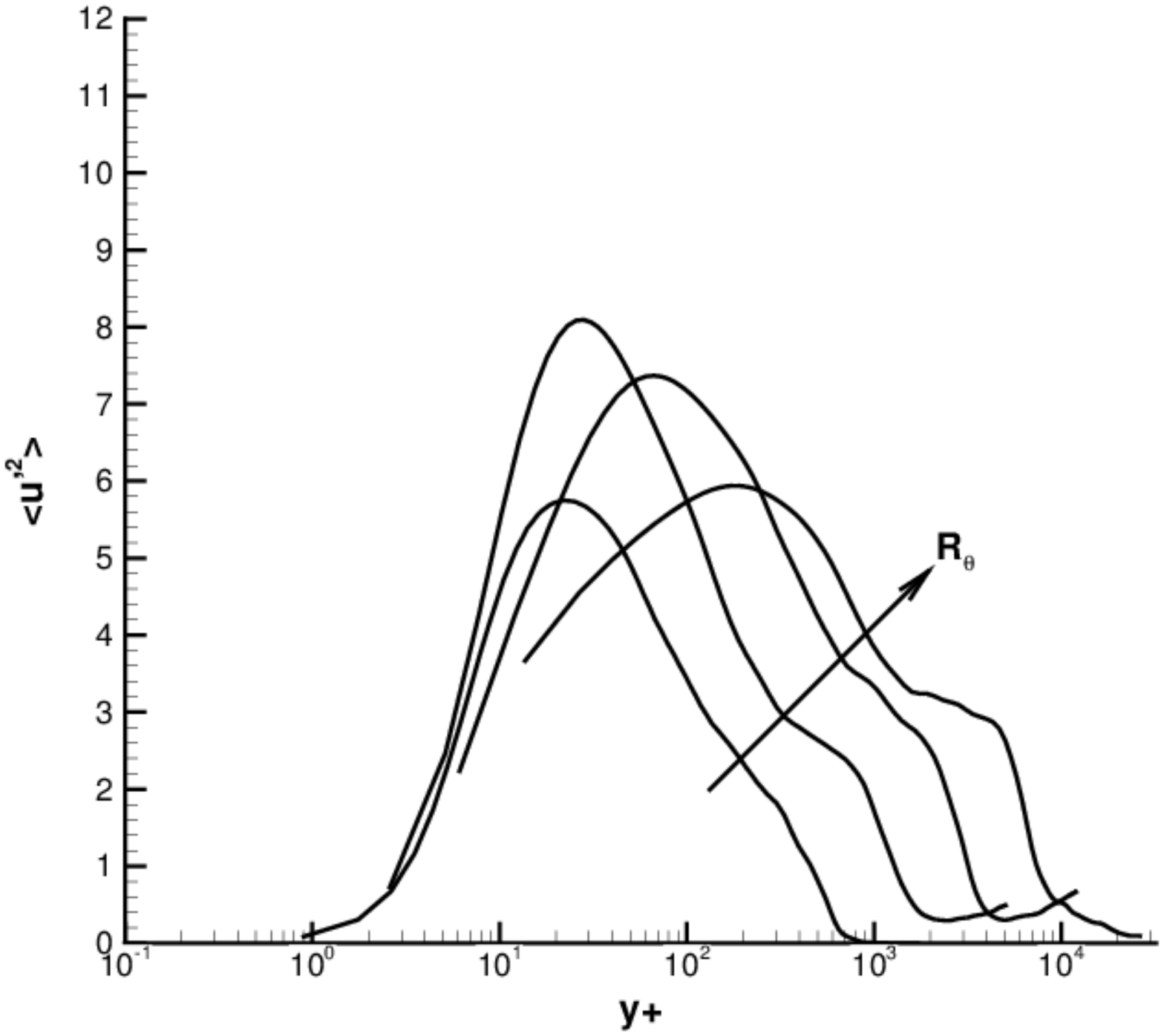}}
}
\end{minipage}
\hfill
\begin{minipage}[t!]{80mm}
\centering{
\subfigure[Wall-normal component]{\includegraphics[width=80mm]{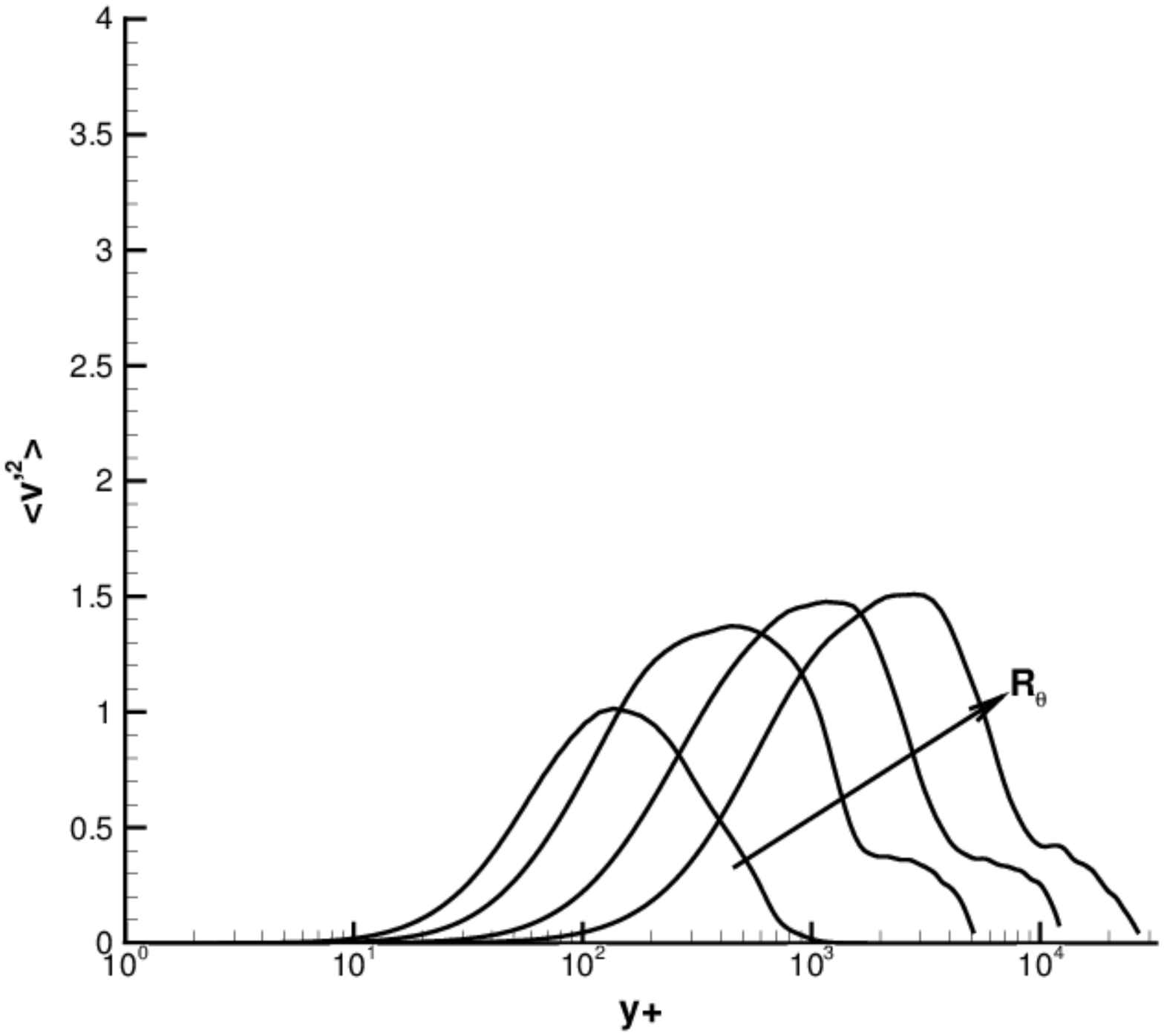}}
}
\end{minipage}
\caption{Streamwise and wall-normal components of Reynolds stress: Variation at different $R_{\theta}$. The arrow indicates direction of increasing momentum thickness Reynolds number.}
\label{ReynoldsStress}
\end{figure}

The trends of Reynolds normal stress components are plotted in figure \ref{ReynoldsStress}. It is generally assumed that if the mean flow near the wall scales on inner coordinates, then so too must the Reynolds stresses, which rely on the mean gradients for their energy \cite{tennekes:1972}. Thus, the Reynolds stresses are almost always normalized by inner scales, $u_{\tau}$ and $\nu/u_{\tau}$. Experimental observations of \cite{degraaff:2000}, however, found that the streamwise component of the Reynolds stress scales on mixed coordinates (inner-outer), and the wall-normal component on the inner coordinates. When a hybrid RANS/LES is used for studying turbulent boundary layers, the near-wall behavior of Reynolds stress depends on the underlying RANS model being used. Here, we use SST variant of the $k-\omega$ as RANS model which gives accurate predictions for the mean velocity, but the near-wall peak of the turbulent kinetic energy is underpredicted due to its formulation \citep{durbin:2011}. So, it is not expected to get accurate results for the Reynolds stress components with hybrid RANS/LES which is why no direct comparisons with the experimental data are made here. The trends plotted show that the near-wall peak of $<{u^{\prime}}^{2}>$ is not constant when $R_{\theta}$ is varied and the location of the peak shifts slightly to the right. This is also true for the wall-normal component of the Reynolds stress. These trends are in good agreement with the observations of \cite{degraaff:2000}, but the inner scaling observed on $<{v^{\prime}}^{2}>$ component is not predicted by hybrid RANS/LES. This could be an artifact of the particular choice of the RANS model itself.

\begin{figure}[!htp]
\begin{minipage}[t!]{80mm}
\centering{
\subfigure[Case A: $R_{\theta}=1520$]{\includegraphics[width=80mm]{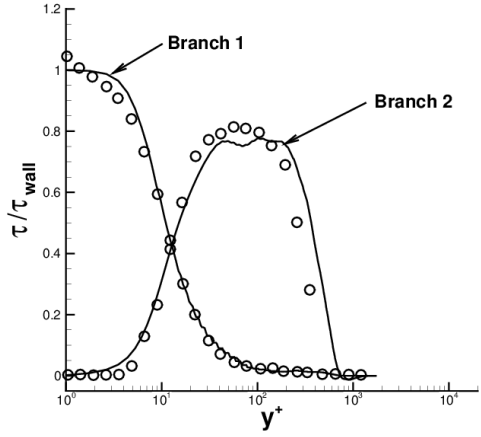}}
}
\end{minipage}
\hfill
\begin{minipage}[t!]{80mm}
\centering{
\subfigure[Case B: $R_{\theta}=5200$]{\includegraphics[width=80mm]{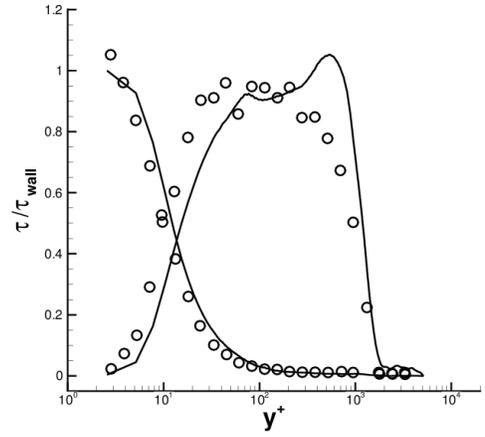}}
}
\end{minipage}
\hfill
\begin{minipage}[t!]{80mm}
\centering{
\subfigure[Case C: $R_{\theta}=13000$]{\includegraphics[width=80mm]{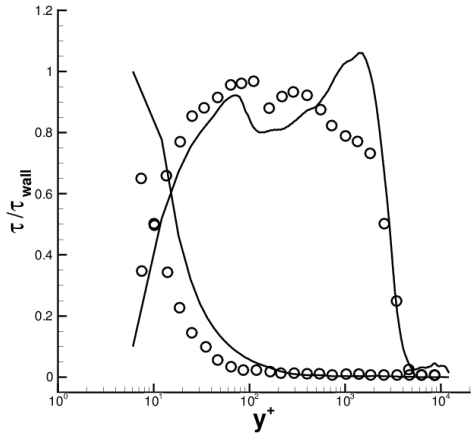}}
}
\end{minipage}
\hfill
\begin{minipage}[t!]{80mm}
\centering{
\subfigure[Case D: $R_{\theta}=31000$]{\includegraphics[width=80mm]{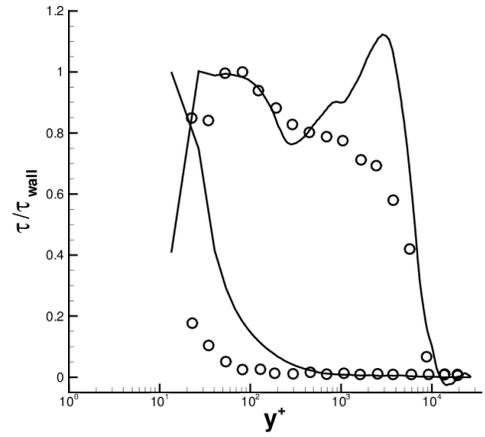}}
}
\end{minipage}
\caption{Shear stress distribution in comparison with the experimental data. Solid lines: IDDES, Symbols: \cite{degraaff:2000}. Branch 1: Viscous shear; Branch 2: Turbulence shear (resolved plus modeled).}
\label{ShearStress}
\end{figure}

Figure \ref{ShearStress} presents shear stress distribution in comparison with the experimental data. It has two branches for each $R_{\theta}$. Branch 1 gives the viscous shear ($\tau_{viscous}$), and Branch 2 gives the turbulence shear ($\tau_{turbulence}$). Note that turbulence shear includes both the resolved and modeled parts as given in the following equation:
\begin{eqnarray}
\tau_{viscous} &=& \mu \frac{dU}{dy} \nonumber \\
\tau_{turbulence} &=& \mu_{t} \frac{dU}{dy}  - {\rho} <u^{\prime}v^{\prime}>
\end{eqnarray}
where $\mu_{t}$ is the eddy viscosity calculated from the underlying SST $k-\omega$ model \citep{menter:1994}. Close to the wall, viscous shear is dominant. Away from the wall (about $y^{+}>100$), turbulence shear becomes dominant. These two branches cross at $y^{+}\approx 10$ and $\tau/\tau_{wall}\approx 0.5$ except at very high momentum thickness Reynolds number. The agreement with the experimental data is good for the cases with $R_{\theta}$ of 1520 and 5200. At higher Reynolds numbers, there is significant error close to the wall which could be attributed to the near-wall modeling. The "bi-modal" shape of the turbulence shear is, perhaps, due to the switch between RANS and LES. The resolved field is showing larger peak in the turbulence shear than in the experiments. These discrepancies indicate that there is scope for further optimizing the parameters used in the length scales of the IDDES model for high $R_{\theta}$.

\begin{figure}[!htp]
\centering{
\includegraphics[width=120mm]{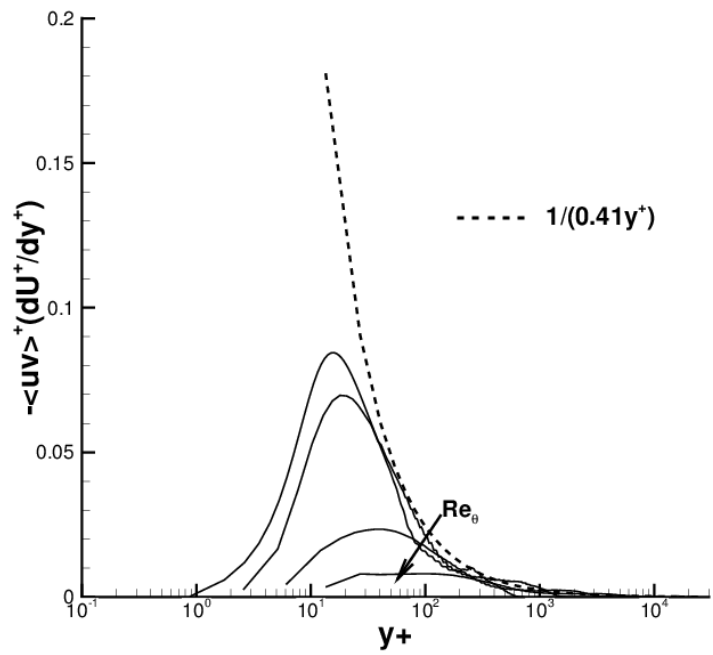}
}
\caption{Reynolds stress production variation with $R_{\theta}$ (calculated from the resolved velocity field).}
\label{ReynoldsStress_production}
\end{figure}

The Reynolds stress production calculated from the resolved field is plotted in figure \ref{ReynoldsStress_production}. It is interesting to see that the peak production decreases with increase in $R_{\theta}$. This means that RANS model is active up to a large $y^{+}$ at higher $R_{\theta}$. The maximum of the production from the resolved field occurs where 
\begin{equation}
-\rho<u^{\prime}v^{\prime}>=\mu(\frac{dU}{dy})
\end{equation}
This corresponds to a location, $y^{+} \approx 15$, where Branch 1 and Branch 2 meet in the shear stress distribution plotted in figure \ref{ShearStress}. Assuming that the total shear stress is approximately constant and equal to $\tau_{w}$ in the log region implies that the production should lie along the line 
\begin{equation}
-<u^{\prime}v^{\prime}>^{+}\left(\frac{dU^{+}}{dy^{+}}\right)=\frac{1}{\kappa y^{+}}.
\end{equation}
This line is also plotted in figure \ref{ReynoldsStress_production} with $\kappa=0.41$. In fact, this is the assumption used in deriving the model constants for the RANS modeling. At higher momentum thickness Reynolds number, log region corresponds to only a small range of $y^{+}$ in our simulations. It is, perhaps, possible to vary the grid resolution to increase the extent of the LES region, but that has not been tried out in this work.

\begin{figure}[!htp]
\centering{
\includegraphics[width=120mm]{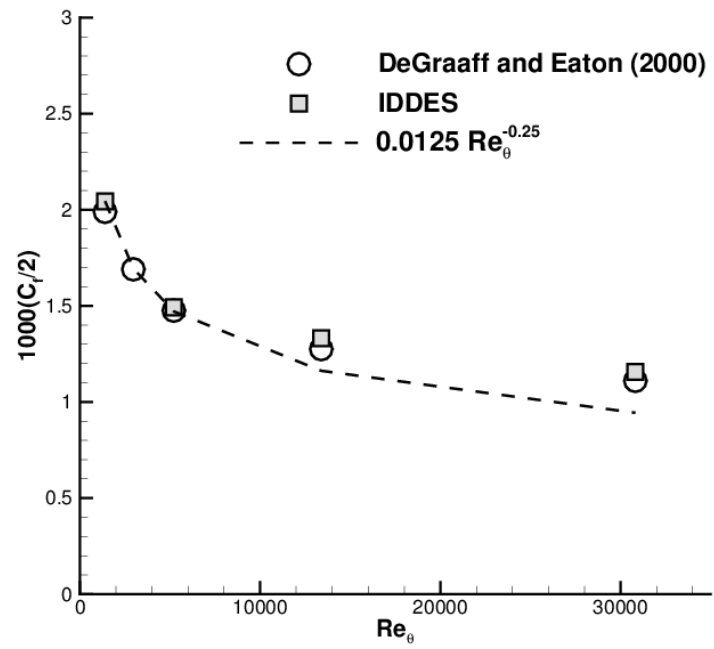}
}
\caption{Skin friction variation with $R_{\theta}$. The dashed line is plotted using a correlation based on the $\frac{1}{7}^{th}$ power law.}
\label{Cf}
\end{figure}

\begin{figure}[!htp]
\centering{
\includegraphics[width=120mm]{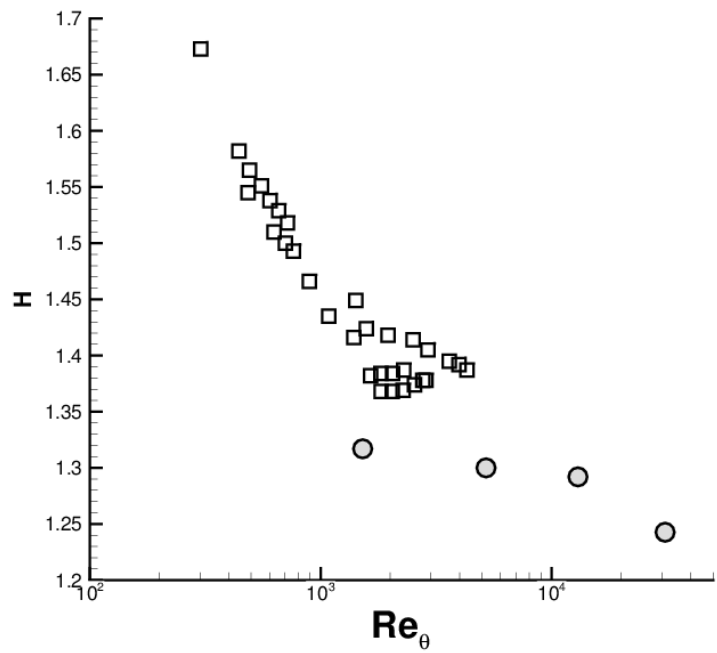}
}
\caption{Shape factor, H, as a function of $R_{\theta}$. Filled symbols are IDDES, Open symbols are from DNS data of \cite{schlatter:2010}.}
\label{ShapeFactor}
\end{figure}

\begin{figure}[!htp]
\centering{
\includegraphics[width=120mm]{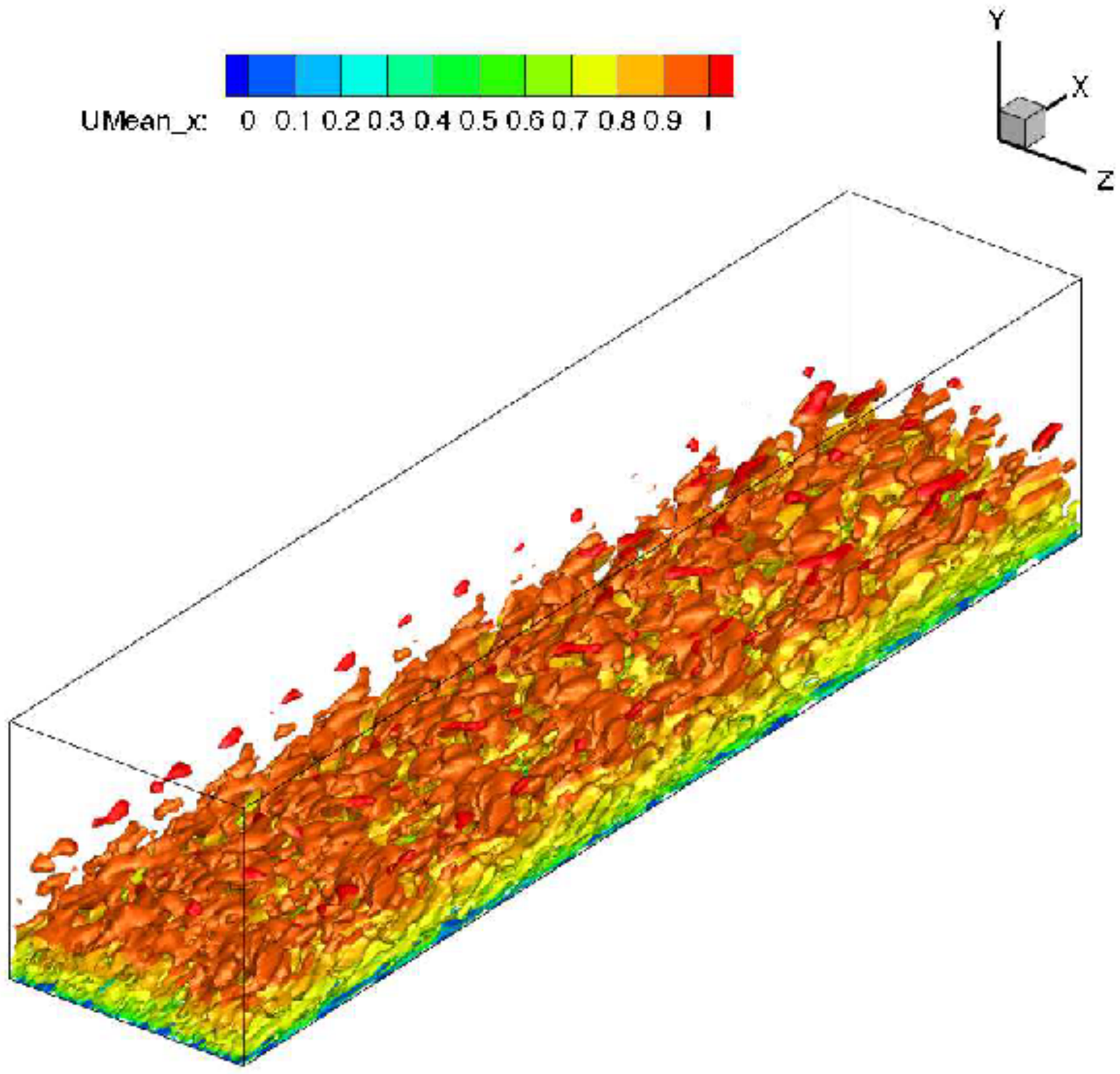}
}
\caption{Vortical structures identified using isosurfaces of Q-criterion at $R_{\theta}=1520$.}
\label{Q-iso1}
\end{figure}

\begin{figure}[!htp]
\centering{
\includegraphics[width=120mm]{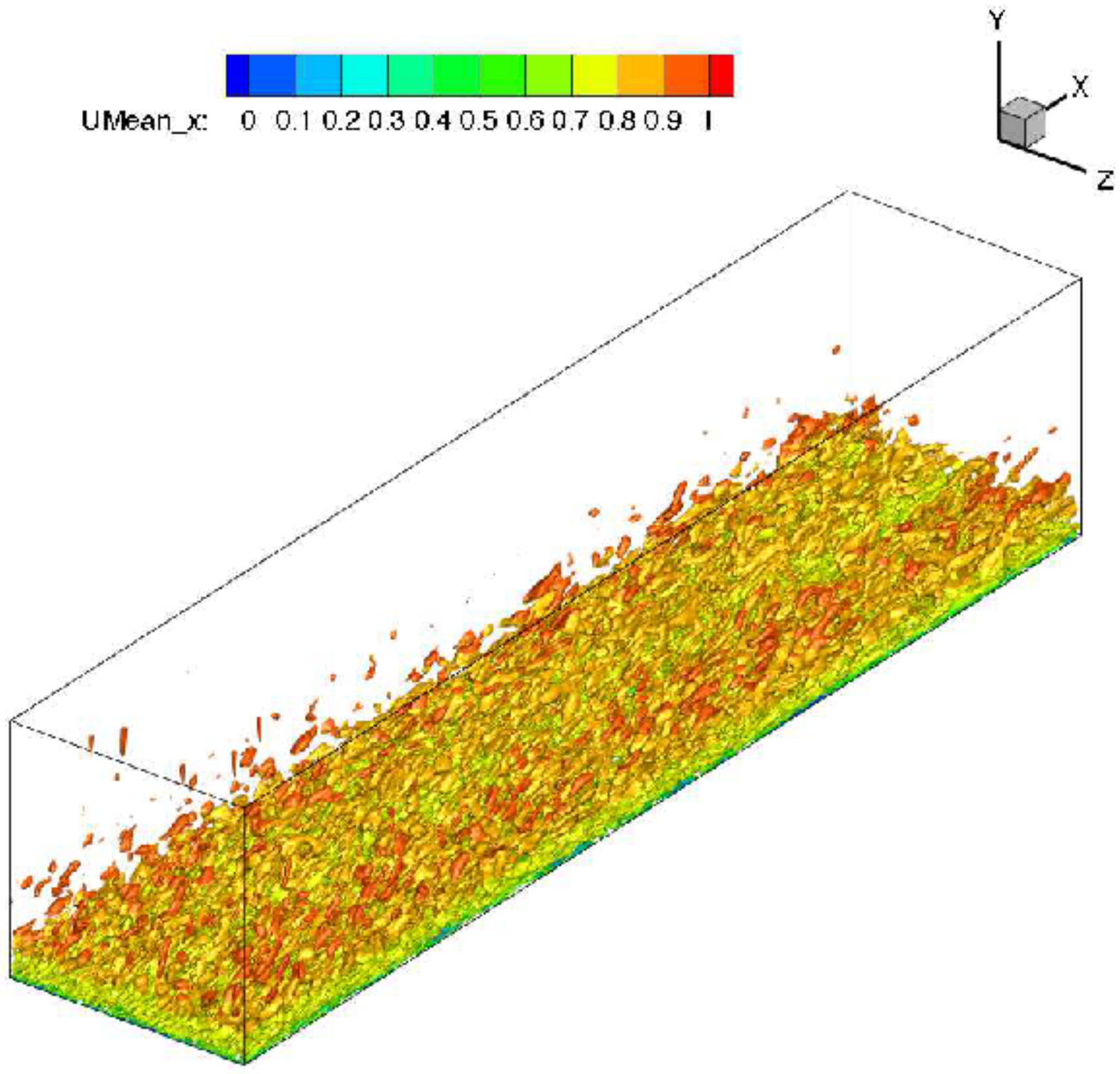}
}
\caption{Vortical structures identified using isosurfaces of Q-criterion at $R_{\theta}=31000$.}
\label{Q-iso2}
\end{figure}

Skin friction is the most important global quantity of interest to the designers. It is plotted in figure \ref{Cf} along with the experimental data and a data correlation from the 1/7th power law. IDDES shows good agreement with the data at all Reynolds numbers considered. Shape factor, $H=\delta^{*}/\theta$ is plotted in figure \ref{ShapeFactor} in comparison with the DNS data of \cite{schlatter:2010}. The trends are predicted well for this quantity. Vortical structures as identified with the Q-criterion are plotted at the lowest and the highest momentum thickness Reynolds numbers in figures \ref{Q-iso1}-\ref{Q-iso2}. As expected, vortical structures resolved at $R_{\theta}=31000$ are much finer scales compared with that at $R_{\theta}=1520$.

\section{Conclusions}
A hybrid RANS/LES framework based on $k-\omega$ SST based IDDES model and a variant of recycling and rescaling method of generating inflow turbulence is presented. Using this framework, numerical simulations of spatially developing flat plate turbulent boundary layers were performed up to $R_{\theta}=31000$. The following are the principal conclusions drawn from this work:
\begin{itemize}
\item IDDES predicted the global quantities such as mean velocity and skin friction accurately in comparison with the experimental data.
\item The detailed shear stress distribution indicate scope for improving the near-wall modeling at high $R_{\theta}$ . It is necessary to further optimize the parameters used in the model for a wide range of Reynolds numbers.
\item Vortical structures show expected behavior at large $R_{\theta}$.
\end{itemize}
This work demonstrates the applicability of IDDES to predict global quantities at high Reynolds numbers without having to use very fine grid resolution. Future work will involve investigating the spatially developing boundary layers under the influence of pressure gradients. The framework presented in this paper will be useful in investigating such benchmark problems in more detail and possibly improving the closure models.

\bibliographystyle{elsarticle-harv}

\bibliography{arxiv_IDDES_TBL}

\end{document}